\documentclass[twocolumn]{aastex62}

\newcommand{\chn}{{\it Chandra}}

\graphicspath{{./}{figures/}}

\shorttitle{Quasar Jets with LOFAR}
\shortauthors{Harris D. E. et al.}

\begin{document}

\title{LOFAR observations of 4C+19.44. \\ On the discovery of low frequency spectral curvature in relativistic jet knots.}

\correspondingauthor{Francesco Massaro}
\email{f.massaro@unito.it}

\author{D. E. Harris}
\affil{Harvard - Smithsonian Astrophysical Observatory, 60 Garden Street, Cambridge, MA 02138.}

\author{J. Mold\'{o}n}
\affiliation{Jodrell Bank Centre for Astrophysics, Alan Turing Building, The University of Manchester, Oxford Road, Manchester, M13 9PL, UK}

\author{J. R. R. Oonk}
\affiliation{ASTRON, the Netherlands Institute for Radio Astronomy, Oude Hoogeveensedijk 4, 7991 PD, Dwingeloo, The Netherlands.}
\affiliation{Leiden Observatory, Leiden University, P.O. Box 9513, NL-2300 RA Leiden, The Netherlands.}
\affiliation{SURFsara, P.O. Box 94613, 1090 GP Amsterdam, The Netherlands.}

\author{F. Massaro}
\affiliation{Dipartimento di Fisica, Universit\`a degli Studi di Torino (UniTO), via Pietro Giuria 1, I-10125 Torino, Italy.}
\affiliation{Istituto Nazionale di Fisica Nucleare, Sezione di Torino, via Pietro Giuria 1, I-10125 Torino, Italy.}
\affiliation{INAF-Osservatorio Astrofisico di Torino, via Osservatorio 20, 10025 Pino Torinese, Italy.}
\affiliation{Consorzio Interuniversitario per la Fisica Spaziale (CIFS), via Pietro Giuria 1, I-10125, Torino, Italy.}

\author{A. Paggi}
\affiliation{Dipartimento di Fisica, Universit\`a degli Studi di Torino (UniTO), via Pietro Giuria 1, I-10125 Torino, Italy.}
\affiliation{Istituto Nazionale di Fisica Nucleare, Sezione di Torino, via Pietro Giuria 1, I-10125 Torino, Italy.}

\author{A. Deller}
\affiliation{Swinburne University of Technology, PO Box 218, Hawthorn VIC 3122 Australia}

\author{L. Godfrey}
\affiliation{ASTRON, the Netherlands Institute for Radio Astronomy, Oude Hoogeveensedijk 4, 7991 PD, Dwingeloo, The Netherlands.}

\author{R. Morganti}
\affiliation{ASTRON, the Netherlands Institute for Radio Astronomy, Oude Hoogeveensedijk 4, 7991 PD, Dwingeloo, The Netherlands.}
\affiliation{Kapteyn Astronomical Institute, University of Groningen, P.O. Box 800, 9700 AV Groningen, The Netherlands.}

\author{S. G. Jorstad}
\affiliation{Institute for Astrophysical Research, Boston University, 725 Commonwealth Avenue, Boston, MA 02215, USA.}
\affiliation{Astronomical Institute, St. Petersburg University, Universitetskij Pr. 28, Petrodvorets, 198504 St. Petersburg, Russia}

\begin{abstract}
We present the first LOFAR observations of the radio jet in the quasar 4C+19.44 (a.k.a. PKS 1354+19) obtained with the long baselines. The achieved resolution is very well matched to that of archival Jansky Very Large Array (JVLA) observations at higher radio frequencies as well as the archival X-ray images obtained with {\it Chandra}. We found that, for several knots along the jet, the radio flux densities measured at hundreds of MHz lie well below the values estimated by extrapolating the GHz spectra. This clearly indicates the presence of spectral curvature. Radio spectral curvature has been already observed in different source classes and/or extended radio structures and it has been often interpreted as due to intrinsic processes, as a curved particle energy distribution, rather than absorption mechanisms ({ Razin-Tsytovich} effect, free-free or synchrotron self absorption to name a few). Here we discuss our results according to the scenario where particles undergo stochastic acceleration mechanisms also in quasar jet knots.
\end{abstract}


\section{Introduction}
\label{sec:intro}
Since the early 1960s, spectral curvature at MHz frequencies was observed in extragalactic radio sources \citep{howard65a}. There are many examples of convex radio spectra in the literature, in particular for unresolved extragalactic sources \citep[see e.g.,][to name a few]{scheuer68,laing80,landau86,jackson01}. Moreover, in the last two decades radio spectral curvature also appeared to be a common feature being observed in several source classes. Hotspots in Cygnus A show a turnover at low radio frequencies (Carilli et al. 1991), a result confirmed by recent observations at hundreds of MHz \citep{mckean16} as well as those of PKS 1421-490 \citep{godfrey09}. Other examples are radio spectra of extended structures in radio galaxies \citep[see e.g.,][]{blundell00,lazio06,hardcastle01,godfrey09,lobes,duffy12} as well as in young radio sources \citep[see e.g,][]{fanti95,dallacasa00,tingay03}. 

Synchrotron emission dominates the radio sky at MHz and GHz frequencies and thus the presence of a low frequency turnover was originally attributed to radiative losses (i.e. synchrotron aging) and/or synchrotron self absorption \citep[see e.g.,][for a recent review]{duffy12}. Other sources of absorption,  that are mainly related to plasma effects, such as external and/or internal free-free absorption, { Razin-Tsytovich} effect and Compton scattering, were also invoked to interpret the observed radio spectral curvature but none of them appeared to be a viable model to describe MHz-to-GHz spectra of both compact radio sources \citep{howard65b,tingay03} and extended structures of radio galaxies \citep{blundell00}. Two additional scenarios were also recently proposed to interpret the low radio frequency flattening of the radio spectra in the hotspots of PKS 1421-490, the first linked to jet energy dissipation and a second related to a transition between two distinct acceleration mechanisms \citep{godfrey09}.  

We are now living in the ``Golden Age'' for radio astronomy at low frequency (i.e. below $\sim$1GHz) and the advent of Low-Frequency Array (LOFAR) allows us to investigate radio spectral curvature with unprecedented sensitivity and resolution. Recent LOFAR observations of the hotspots in Cygnus A not only confirmed the presence of a low frequency turnover, but accurately quantified the spectral shape at and below the turnover frequency. The low frequency spectral curvature was shown to be too rapid to be the result of a cutoff in the electron energy distribution (assuming a standard synchrotron kernel with isotropic pitch angle distribution and isotropic magnetic field distribution), necessitating some form of absorption, or a non-standard synchrotron kernel to explain the observed low-frequency spectra. The model parameters required for both synchrotron self-absorption and free-free absorption are problematic, and the cause of spectral curvature in the hotspot of Cygnus A remains a mystery \citep{mckean16}.

The knowledge of the spectral shape at low (i.e. hundreds of MHz) radio frequencies permits us to determine when a power-law spectrum can be properly extrapolated below the GHz regime. This has several important consequences potentially affecting estimates of source parameters. Quantities such as the equipartition field $B_{eq}$, the total energy $E_{tot}$, and the non-thermal pressure will be more accurately estimated by the determination of the low frequency spectra. To compute the minimum energy and pressure in non-thermal plasmas, the most conservative approach is to integrate the electron spectrum only over those energies producing observable emission. However, because of the steepness of the electron spectrum, it is often the case that most of the total energy resides in the lowest energies so when these are ignored, we introduce significant uncertainties into our estimates of $B_{eq}$, $E_{tot}$, pressure, etc. Obtaining radio spectra to low frequencies will significantly reduce these uncertainties. In addition radio observations at tens of MHz could be the key to shed a light on the nature of the X-ray emission in quasar jets \citep[see e.g.,][]{harris02,harris06,worrall09}. According to the well-entertained IC/CMB scenario \citep{hoyle65,bergamini67,harris79,schwartz00,chartas00}, where X-ray emission of jet knots arises from inverse Compton scattering between particles (i.e., electrons) accelerated in the jet and seed photons from the Cosmic Microwave background (CMB), the spectral behavior of particles responsible for the low radio frequency emission is directly observable in the X-rays.

Moreover, the origin of X-ray emission from powerful radio sources, such as quasars, is still debated with synchrotron emission and IC/CMB radiation competing. In particular, Harris and Krawczynski (2006) summarized several arguments against the IC/CMB model for the X-ray emission of jet knots but a conclusive answer is still unknown. Recently Cara et al. (2013) showed that polarized optical emission discovered in the quasar jet of PKS\,1136-135 rules out the IC/CMB scenario in which radiation is expected to be unpolarized. This effect has been also detected in Pictor A \citep[see e.g.,][for more details]{gentry15} and 3CR\,111 \citep{clautice16}. Furthermore, the lack of gamma-ray emission observed in the nearby jets of 3C\,273 and PKS\,0637-752 favored the synchrotron interpretation for the X-ray emission with respect to the competing IC/CMB scenario \citep{meyer14,meyer15,meyer17}, at least in low redshift sources (i.e., $z$ lower than $\sim$0.5).

Radio observations in the MHz energy range will be crucial to improve our estimates of source parameters as $B_{eq}$, $E_{tot}$ even if we will not be able to investigate completely the low energy tail of the emitting particles.

Motivated by the importance of obtaining low radio frequency observations for extragalactic radio sources here we present the results of LOFAR observations of a quasar jet in 4C +19.44. 
By using the international baselines of the { LOFAR} \citep{haarlem13,moldon15,varenius16,morabito16}  we have been able to reach a resolution at $\sim 0.4$ arcsec at 150~MHz, thus matching the resolution of available JVLA observations at 5~GHz \citep{harris17}. These observations allowed us to resolve, spatially, the quasar jet in 4C+19.44, highlighting the knotty structure at these wavelengths.

The LOFAR images of 4C +19.44 presented here are among the best obtained to date at low radio frequency with sub-arcsecond resolution \citep[see][for other examples]{varenius16,olivencia18} (and arcsec resolution at even lower frequencies, \cite{morabito16}, and the best in term of detailed resolved jet structure. 

For the first time we report the discovery of radio spectral curvature, at $\sim$100MHz, for jet knots in a quasar, in addition to the previously mentioned classes of extragalactic sources. Implications of our results are discussed in the framework of the well-known stochastic acceleration scenario that lead to log-parabolic (i.e. log-normal) spectral energy distributions \citep[see e.g.][]{kardashev62,massaro04,tramacere11}. 

For our analysis, we use cgs units, unless stated otherwise, and we adopt $h$=$H_0/(100\,\mathrm{km\,s^{-1}\,Mpc^{-1}})$=0.71, $\Omega_M$=0.27 and $\Omega_\lambda$=0.73, so that at the source redshift of 0.719 \citep{steidel91}, 1\arcsec corresponds to 7.2 kpc. Spectral indices, $\alpha$, are defined by flux density $S_{\nu} \propto \nu^{-\alpha}$.
\begin{figure*}[!th]
\begin{center}
\includegraphics[height=9.5cm,angle=0]{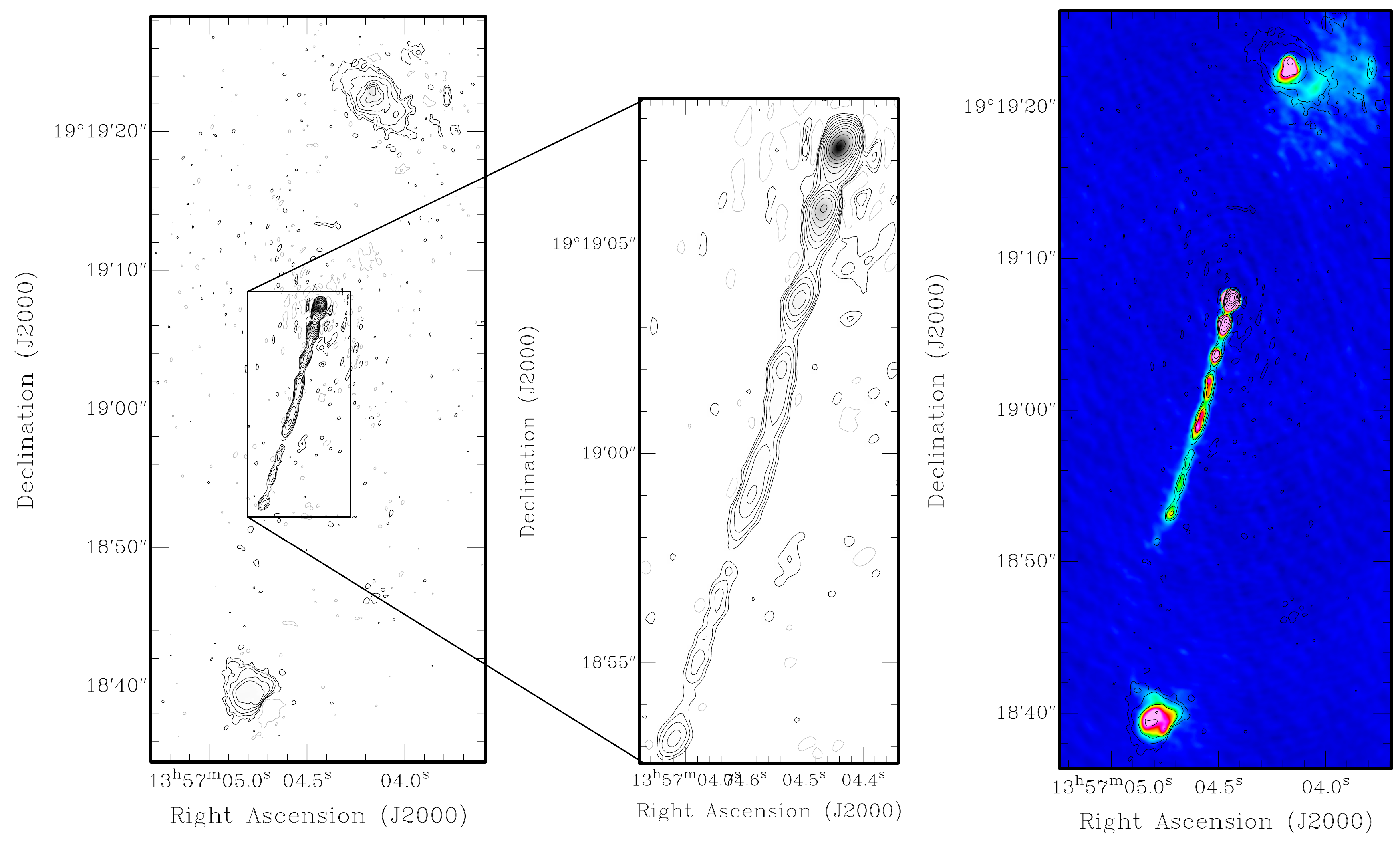}
\caption{{ Left and center:} LOFAR image of 4C+19.44 with zoom-in (center) of the jet. The contour levels are -0.0012, 0.0012 to 1.5 in multiple of 2.  { Right:} The JVLA radio image of 4C+19.44 in C band with LOFAR radio contours overlaid. 
The image clearly shows how the radio structure observed at 159 MHz matches the morphology at GHz frequencies. Extended structures in the northern lobe are detected at larger significance in the JVLA image, but jet knots are detected in both images as we expected when selecting the target to carry out our experiment.  
}
\label{fig:jet1}
\end{center}
\end{figure*}

\section{4C+19.44: a case study}
\label{sec:target}
To select a suitable target for carrying out LOFAR observations, we reviewed the lists of quasars in the XJET webpage\footnote{http://hea-www.harvard.edu/XJET/} \citep{xjetweb,xjet} and chose candidates according to the following criteria:
\begin{enumerate}
\item a redshift value lower than 1.5. This criterion was chosen to avoid higher redshift quasars for which the $(1+z)^4$ increase in the CMB energy density precludes the necessity for significant values of $\delta$ and/or $\Gamma$. Nevertheless surface brightness also scales as $(1+z)^4$ thus targets lying at higher redshifts are not ideal to investigate extended radio structures.
\item Declination grater than 10\degr\ (i.e., in the northern hemisphere) and suitable for obtaining reasonable $u,v$ coverage with LOFAR. This guarantees the high resolution at low radio frequencies necessary to compare the radio image with those available in the Jansky Very Large Array (JVLA) and in the \chn\ archives.
\item Radio and X-ray detections along the jet at distances greater than 2\arcsec, with preference for multiple knot X-ray detections. This will favor future investigations of the IC/CMB scenario.
\end{enumerate}

The only candidate matching all above criteria is 4C+19.44 at $z$=0.719.  This object also has a very good coverage in the L, C, and U radio bands in the JVLA archive as well as relatively long \chn\ observations with exposures of more than $\sim$200 ksec in total \citep{harris17}. Such X-ray data constitute a perfect set of observations for further investigation on the IC/CMB scenario. { The radio structure of 4C+19.44 is composed of a jet with 11 knots extending in the southern direction and terminating with an hotspot lying at about 26\arcsec from the core, while a single lobe is present in the northern side.}
\begin{figure*}[!th]
\includegraphics[width=18cm,angle=0]{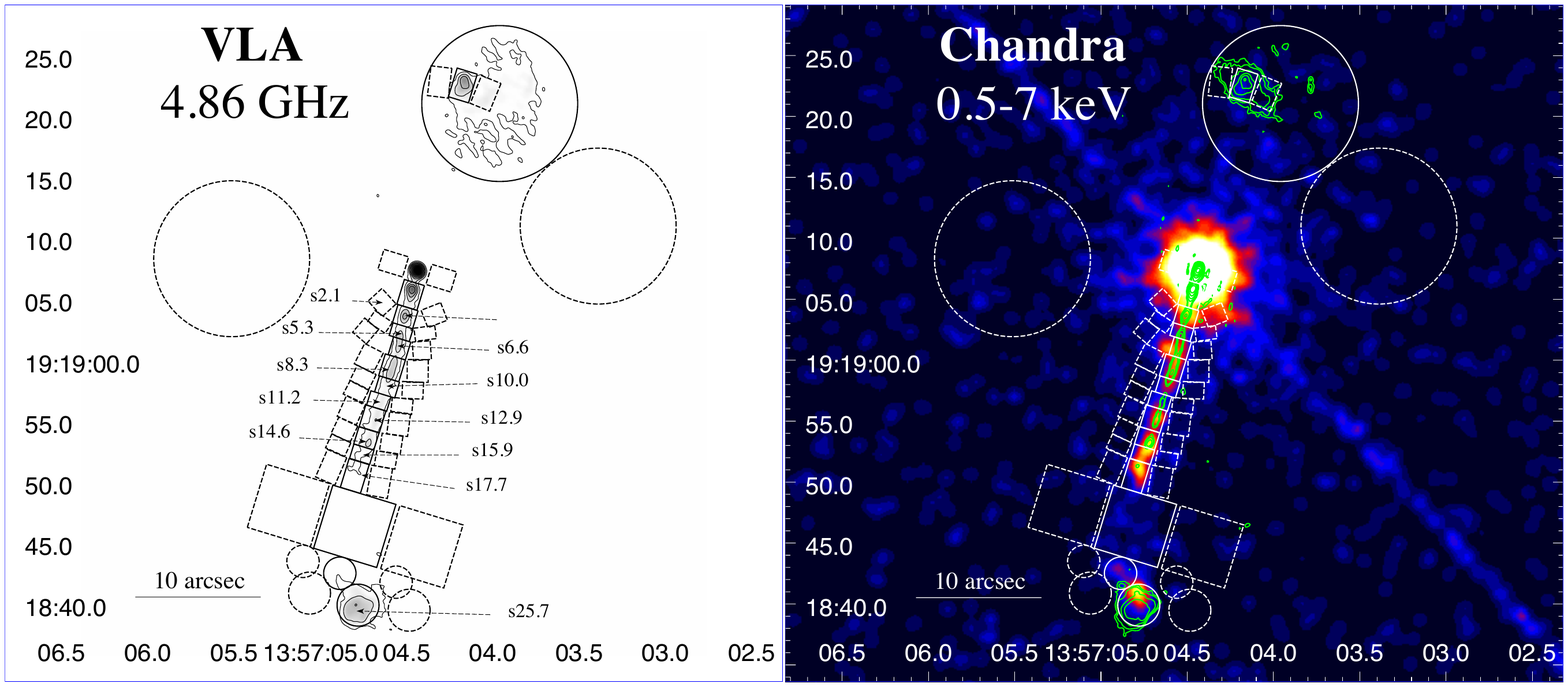}
\caption{JVLA 5~GHz image (left) and {\it Chandra} image (right) with superposed the LOFAR contours (green) illustrating the clear similarity of their spatial resolution. The black boxes correspond to the regions, { of all 11 jet knots and the southern hotspot,} used for extracting radio fluxes (see Table in the Appendix) and deriving the spectral properties, see Fig.~\ref{fig:jet3}. Regions marked with dashed lines are those used as background in the previous X-ray analyses, shown here for completeness \citep{xjet,harris17}.}
\label{fig:jet2}
\end{figure*}

\section{Results}

\subsection{The LOFAR image of 4C+19.44}
\label{sec:LOFARimage}
4C+19.44 has been observed with LOFAR at 150 MHz on 14 and 15 May 2014. A detailed description of the observations is give in Appendix 1. 

The full resolution image has a resolution of $0.44 \times 0.33$ arcsec in PA $= -36.5^{\circ}$ and a noise level of $\sim 0.34$ mJy/beam. This image was made at a central frequency of 159 MHz using 3 MHz bandwidth. Fig. \ref{fig:jet1} left shows the structure of 4C+19.44 at this resolution with a zoom-in of the jet. The LOFAR image shows for the first time at low frequency all details of the { radio} structure as seen at high frequency (and in the X-ray band) and described in \citep{harris17}: a bright compact core, a prominent straight and knotty jet to the southeast up to 17'' from the core at position angle $\sim 165^\circ$, a southern hotspot with faint diffuse emission, and a diffuse northern lobe with a hotspot.
The morphology of the LOFAR and JVLA 5~GHz image are strikingly similar and the overlay illustrating this similarity is shown in Fig.\ref{fig:jet1} right.  

Interestingly, only the Northern lobe shows differences between LOFAR and JVLA, with a pronounced extension to the E only detected by LOFAR. This is likely due to its steep spectral shape, as often observed in extended radio lobes. More surprising is the extended radio emission arising from the same lobe that is instead visible on the Western side in the JVLA image appearing not so prominent at LOFAR frequencies. If real, this would suggest emission with an unusual spectral index for this extended emission. The lack of such extended emission at low radio frequencies, in the MHz range, is mostly due to the very high spatial resolution of the LOFAR image, and the corresponding limited sensitivity to low surface brightness extension. This does not affect our conclusions about the small-scale knots.


\subsection{The spectra of the knots}
\label{sec:LOFARspectra}

Figure~\ref{fig:jet2} shows the details of the jet structure and the location of the regions corresponding { to 11 jet knots and the southern hotspot} selected to carry out our analysis. These are the same regions adopted in previous analyses \citep{xjet,harris17} labeled using the nomenclature proposed in the XJET database, where each knot name is a combination of one letter indicating the orientation of the radio structure and one number indicating distance from the core in arcseconds. Resolution of both radio images (see Figs. ~\ref{fig:jet1} and \ref{fig:jet2}) clearly matches that of the \chn\ one.

\begin{figure*}[!th]
\includegraphics[height=6.5cm,width=6.5cm,angle=0]{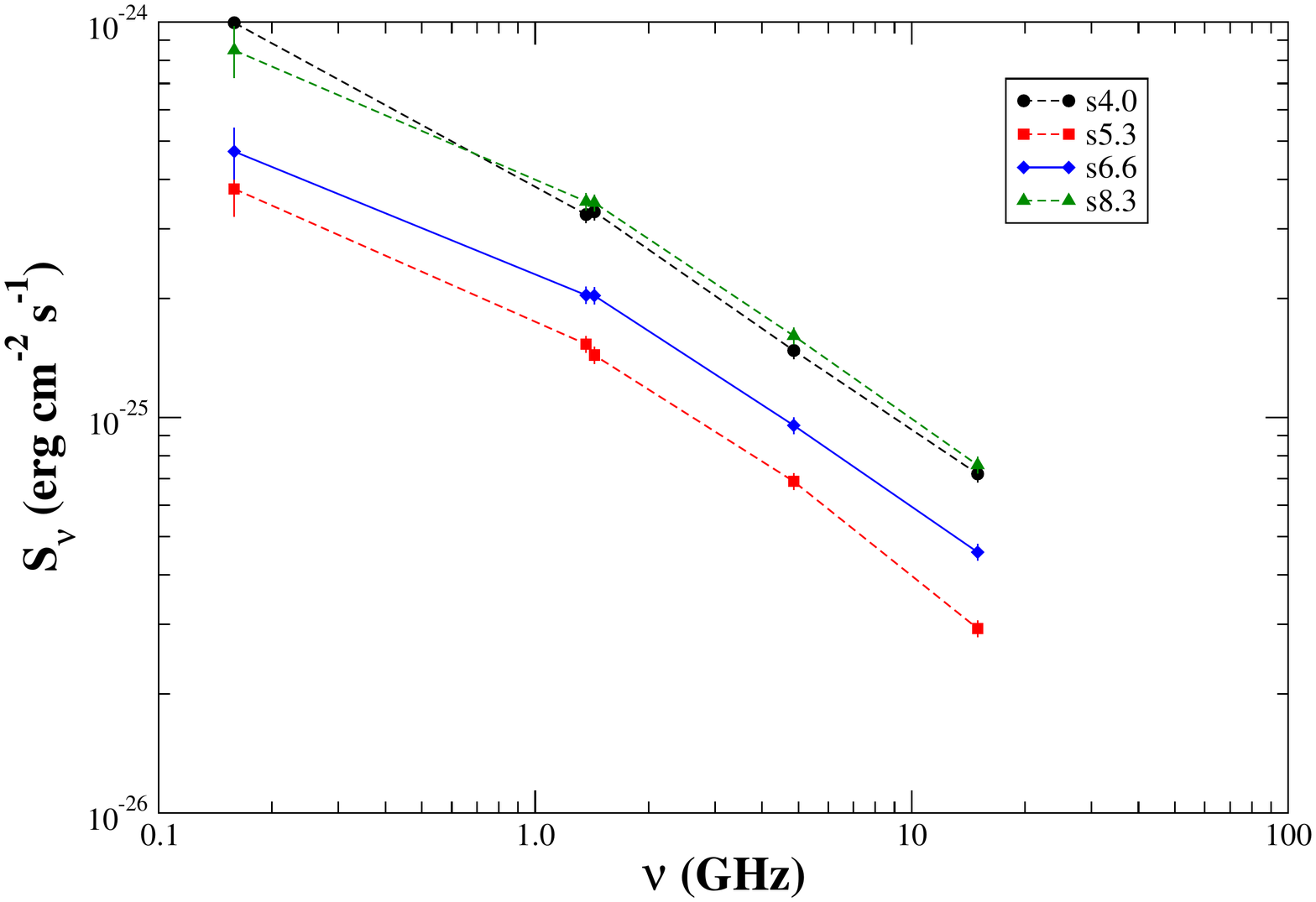}
\includegraphics[height=6.5cm,width=6.5cm,angle=0]{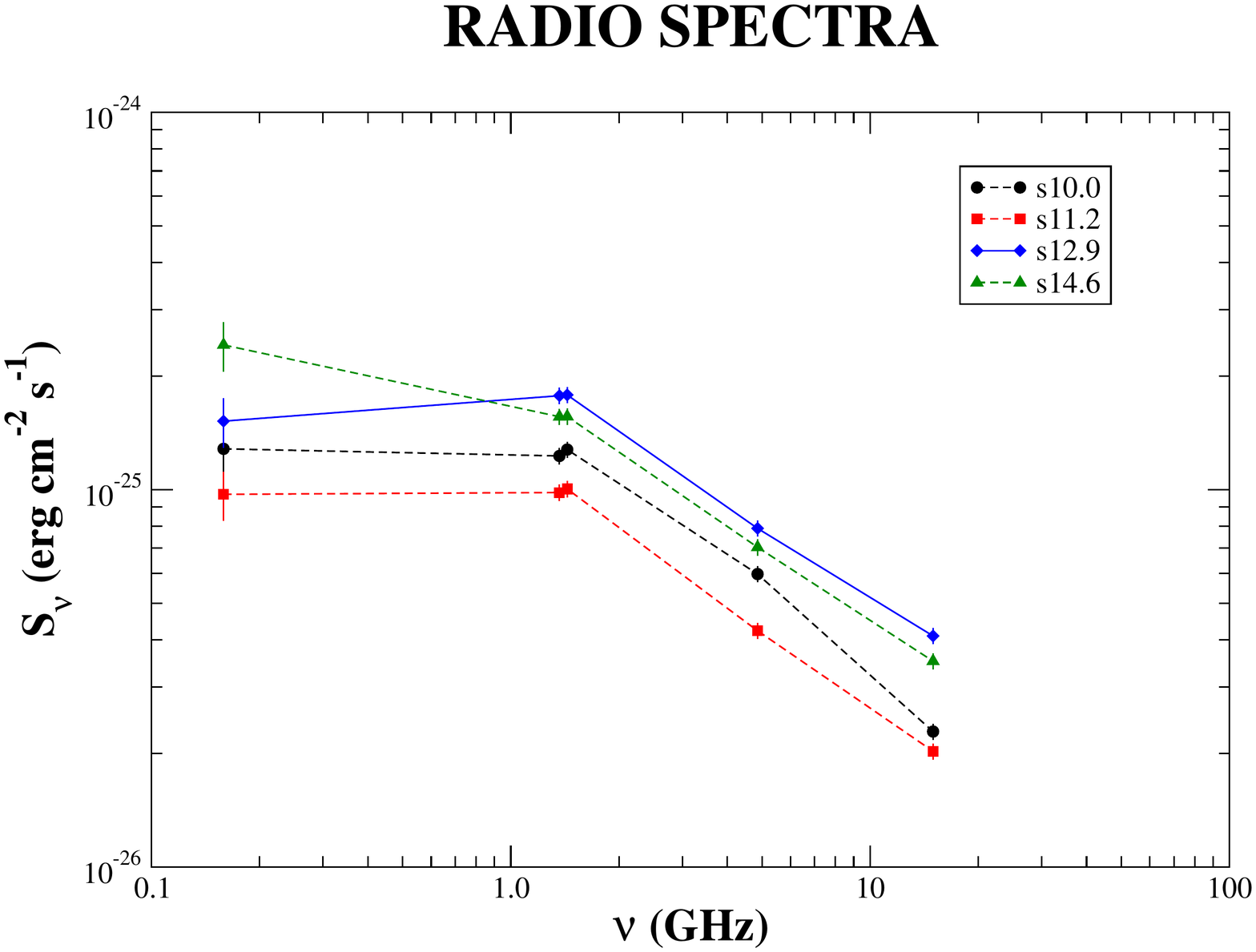}
\includegraphics[height=6.5cm,width=6.5cm,angle=0]{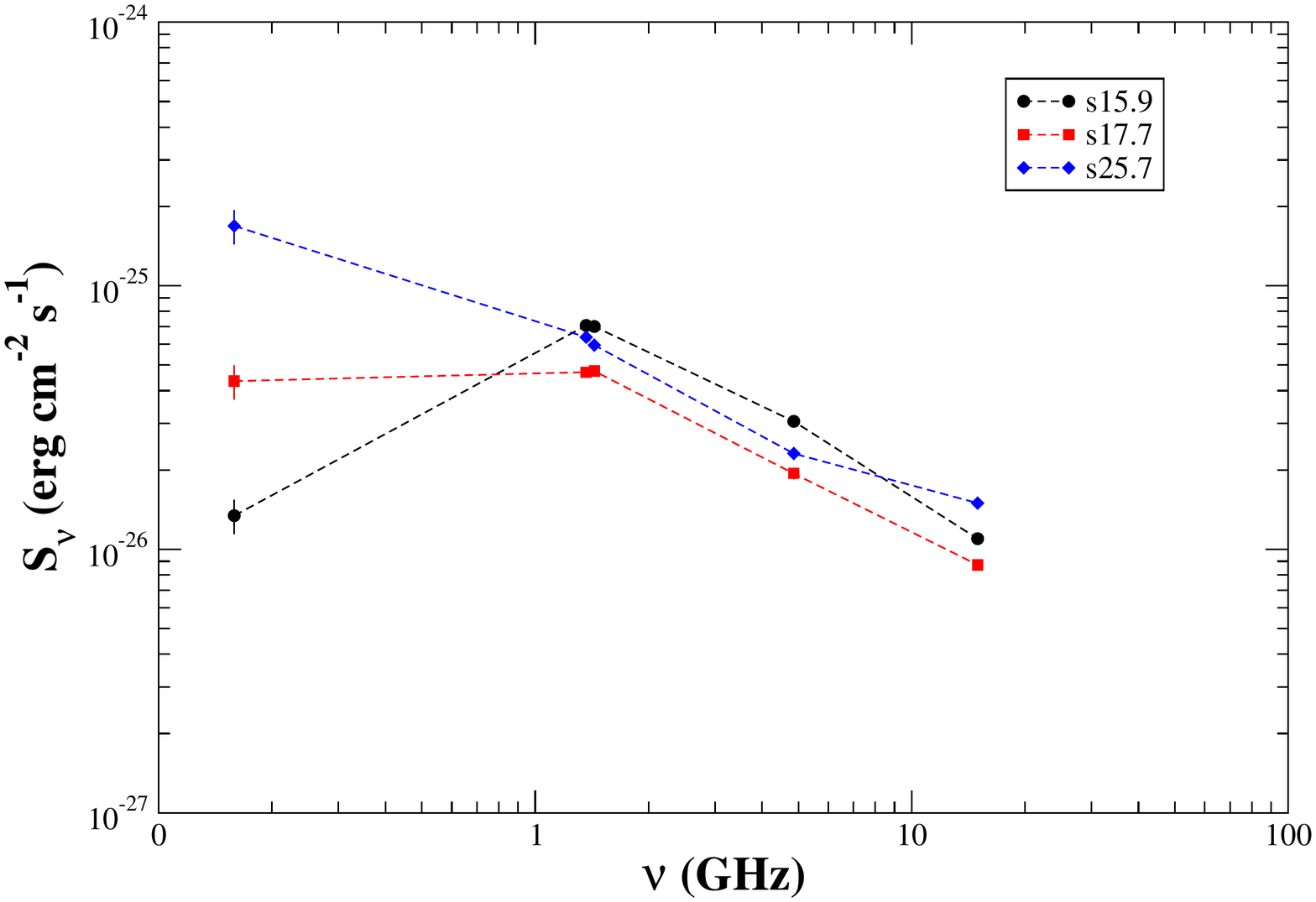}
\caption{{ The radio spectra for 10 knots along the jet labeled following the nomenclature of Fig.~\ref{fig:jet2} and that of the southern hotspot s25.7. We excluded from this figure the knot s2.1 but its radio flux density at low radio frequency is reported in Appendix together with all other radio extended components.}}
\label{fig:jet3}
\end{figure*}

For all jet knots we measured the radio flux density at 159 MHz and we plot it together with those available at higher radio frequencies at few GHz. Flux densities were measured adopting the same regions chosen on the basis of the comparative radio and X-ray analysis \citep[as presented in][]{xjet,harris17}. In particular all boxes shown in  Fig.\ref{fig:jet2}) and marked with dashed lines refer to the background regions used only for extracting the X-ray fluxes and reported here for completeness. Radio flux densities are listed in the Appendix while radio spectra for all knots are shown in Fig.~\ref{fig:jet3}.

It is quite evident that for several jet knots the flux densities measured at 159 MHz lies below the values expected from the extrapolation of the radio spectrum at higher frequencies. Moving toward the most distance knot the effect appears to be more prominent { with the exception of the southern hotspot s25.7}. Radio knots in the jet center show a flat radio spectrum below $\sim$1 GHz and the radio spectrum of the s15.9 knot is clearly inverted. It is worth highlighting that such discrepancy does not appear to be a systematic uncertainty, potentially due to the data reduction processes, since it is not the same for all jet knots. Systematic uncertainties on the radio flux densities have been estimated at 15\% confidence level for the LOFAR data and 5\% for the JVLA data. Furthermore, the reliability of the observed trends was also verified using different sizes of the boxes, indicating that the flux (and therefore the spectral shape) is dominated by the bright knots and not by the underlying diffuse emission.  

\begin{figure}[!th]
\includegraphics[height=8.5cm,width=8.5cm,angle=0]{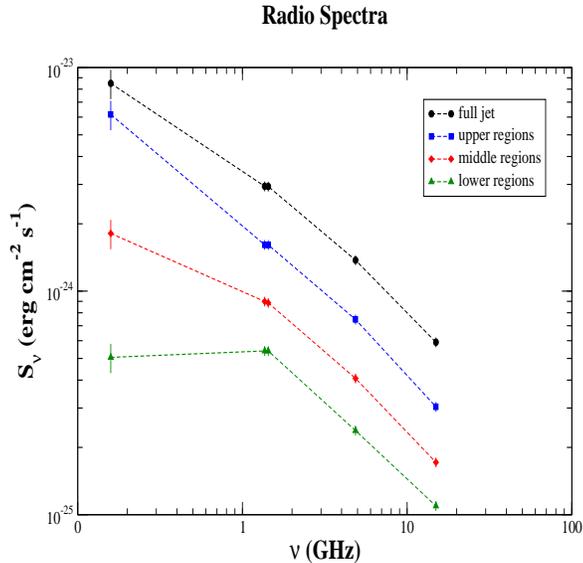}
\caption{The radio spectrum measured over larger regions created merging those selected in previous multi-frequency analyses \citep{xjet} as described in \S~\ref{sec:LOFARspectra}.}
\label{fig:entire}
\end{figure}

We also created three larger regions along the jet, the first including the first two boxes while the other two merging four knot regions each (labeled as upper, middle and lower regions in Fig. ~\ref{fig:entire}). This allowed us to measure, spatially-averaged radio spectra along the radio structure. As shown in Fig.~\ref{fig:entire}, spectral curvature clearly appears also with these larger regions and it seems to be more pronounced moving away from the radio core. 

Since absorption processes (i.e., free-free absorption, { Razin-Tsytovich} effect etc.) as well as radiative losses failed to explain the radio spectral curvature, mostly due to low value of the plasma density estimated in extended structures of radio sources, Duffy \& Blundell (2012) recently proposed to use an intrinsically curved particle energy distribution (PED; i.e., number of particles per unit of volume and their Lorentz factor $\gamma$: $n(\gamma)=\frac{dN}{dVd\gamma}$) in the form of a log-parabola (i.e. log-normal) function to interpret curved radio spectra of radio galaxy lobes. A simple log-parabolic model was even adopted to fit the radio spectra when they were originally observed in compact radio sources \citep{howard65a}. Then, in the last decades log-normal distributions have been extensively used to fit synchrotron spectra of blazars \citep[see e.g.,][]{massaro06,tramacere07,massaro08} as well as those at higher energies of gamma-ray bursts \citep{massaro10,massaro11b}.

The underlying physical explanation for a log-parabolic PED was known and expected by analytical solution of the kinetic equation since the early '60s \citep[see e.g.][]{kardashev62}. In a simple stochastic acceleration mechanism, the low energy turnover is related to the fact that low energy particles gain less energy than high energy ones for each acceleration step \citep{massaro04,massaro06}. On the other hand, high energy particles, having larger Larmor radii when moving in a magnetic field, tend to have lower probability to be accelerated. Both effects produce the mild spectral curvature observed in several source classes. Considering the Fermi acceleration mechanisms \citep{fermi49} and simply assuming that the probability of accelerating particles is not constant leads directly to a log-parabolic PEDs. According to the radio spectra of our 4C+19.44 jet knots we clearly see the effect at low energies but we cannot detect spectral curvature above a few GHz. { We expect that such situation does not occur in the southern hotspot where {\it in situ} particle acceleration mechanism maintain its efficiency \citep[see e.g.,][and references therein]{brunetti03,hardcastle04,cheung05,orienti12,orienti17} and in fact its radio spectral shape does not appear to be curved as shown in Fig~\ref{fig:jet3}.}

It is worth highlighting that the log-parabolic spectrum is defined by only one more parameter with respect to the simple power-law model. According to this scenario we fitted the radio spectra of all jet knots adopting log-parabolic model, tested under the form $S_\nu = S_0\,(\nu/\nu_0)^{-a-b\,log(\nu/\nu_0)}$, where we fixed the value of the spectral curvature $b$ equal to zero above $\nu_0$ to reproduce the high energy tail that is more consistent with a simple power-law spectrum. This spectral shape arises from a log-normal PED described as $n(\gamma)=n_0\,(\gamma/\gamma_0)^{-s-r\,log(\nu/\nu_0)}$ or equivalently defined in terms of the three parameters: the PED peak energy $\gamma_p\,mc^2$, the curvature $r$ and the density $n(\gamma_p)$. Spectral curvature $b$ measured along the jet knots is consistent with the values observed in different source classes ranging between 0.05 and 0.3.
\begin{figure*}[!ht]
\includegraphics[height=6.cm,width=8.5cm,angle=0]{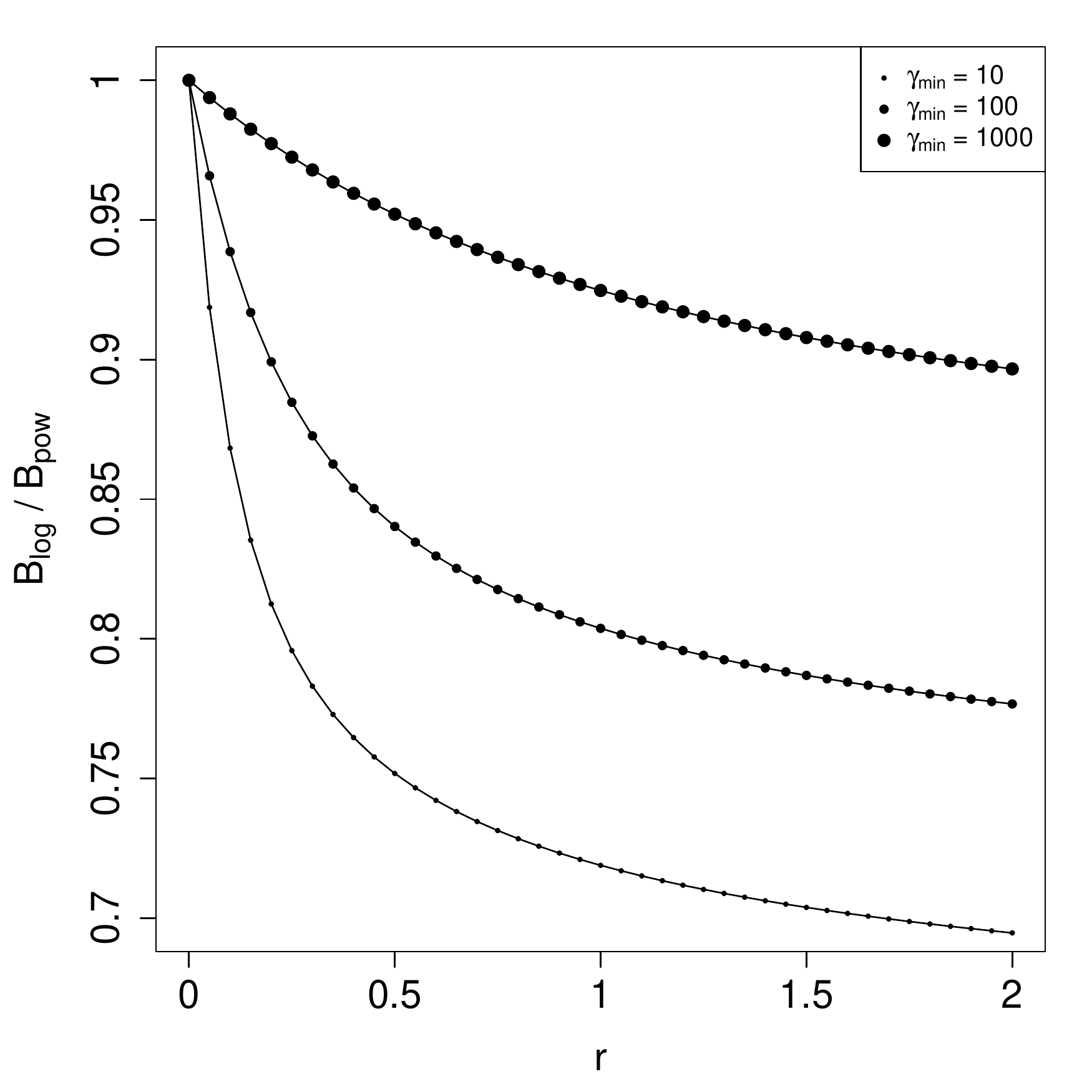}
\includegraphics[height=6.cm,width=8.5cm,angle=0]{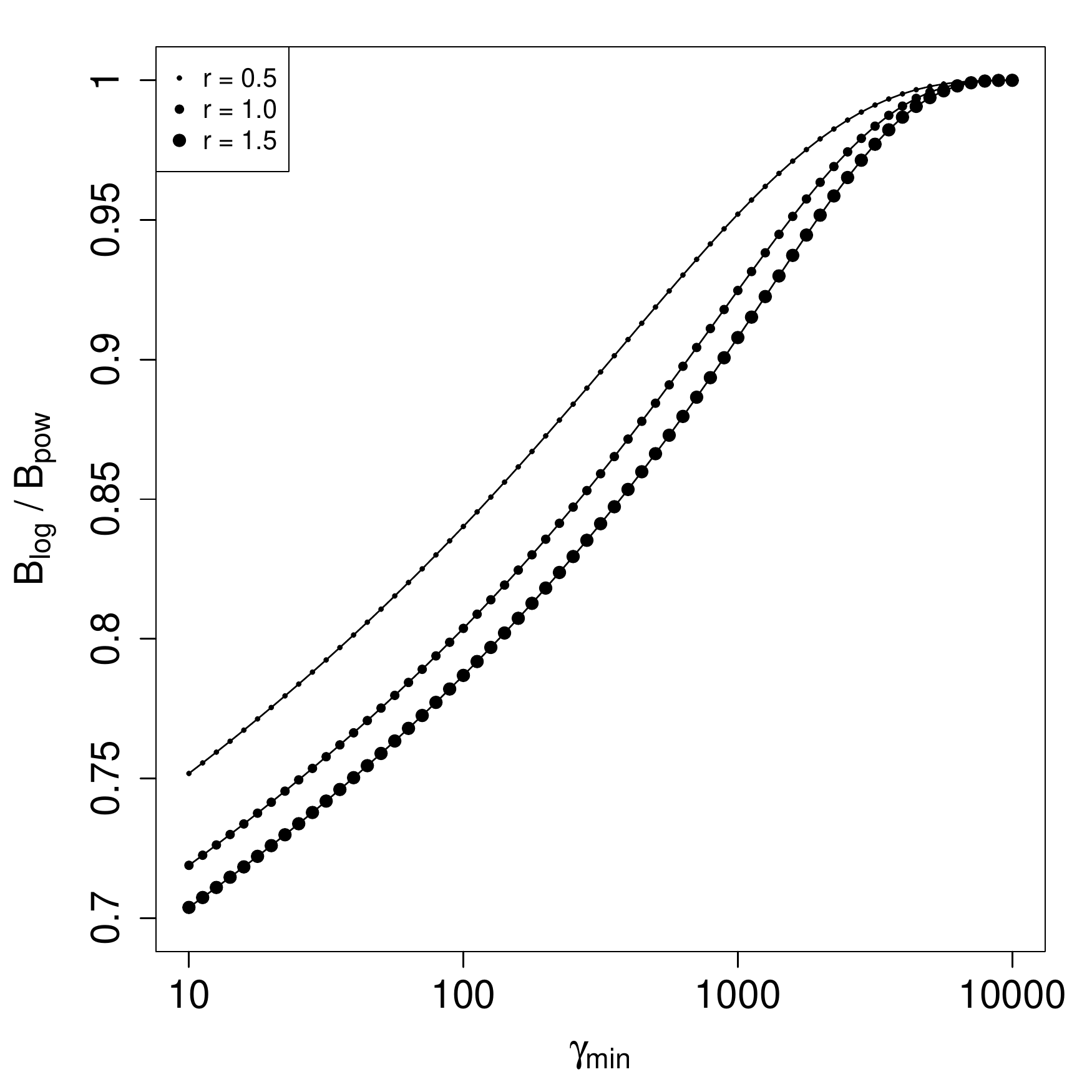}
\caption{The ratio between the magnetic field strength estimated using a simple power-law $vs$ a log-parabolic distribution as function of both the PED curvature $r$ (let panel) and the minimum value of the Lorentz factor for the PED: $\gamma_{min}$ (right panel).}
\label{fig:compare}
\end{figure*}

Finally, we remark that an additional advantage of the log-parabolic particle energy distribution is that the total energy of the emitting electrons $E_{tot}$ is less dependent on the value of the minimum Lorentz factor $\gamma_{min}$ than a simple power-law spectrum.

\section{Estimating Magnetic Field Strengths}
To constrain jet properties, we need to estimate the magnetic field strength. To interpret broad band spectra of extended structures (i.e., jet knots, hotspots, lobes) in radio galaxies two main { assumptions} are generally adopted \citep[see e.g.,][and references therein]{longair11}. The first option is to consider the minimum energy requirements $B_{min}$ having $\frac{\partial}{\partial\,B} (u_B+u_e) =0$, where $u_B=\frac{B^2}{8\pi}$ is the energy density of the magnetic field and $u_e=m_e\,c^2\int_{\gamma_{min}}^{\gamma_{max}}\gamma\,n(\gamma)d\gamma$ that of emitting electrons. The $B_{min}$ estimate is close to the second option that is to assume equipartition condition. According to this assumption the energy density of the magnetic field is the same of that stored in the electrons (i.e., $u_B = u_e$) thus leading to compute $B_{eq}$.

To show how the presence of spectral curvature affects the estimate of the magnetic field strength, we have considered the equipartition condition and calculated the ratio between the equipartition $B_{eq}$ magnetic field computed using a log-parabolic PED and a simple power-law model with the same pivot Lorentz factor $\gamma_0$, the same normalization $n(\gamma_0)$ and the same spectral index $s$.

In Figure~\ref{fig:compare} we show the ratio between the magnetic field strength estimated using a simple power-law $vs$ a log-parabolic distribution as function of both the PED curvature $r$ and the minimum value of the Lorentz factor for the PED: $\gamma_{min}$. It is worth highlighting that estimate of $B_{eq}$ calculated using a power-law model are mainly dependent on $\gamma_{min}$ while in the log-parabolic PEDs on $\gamma_p$. Moreover, differences in the ratio of $B_{eq}$ estimated using the two models are in the range between a few percent and no more than 30\%.

\section{Summary and Conclusions}
We presented here one of the first LOFAR image, obtained with the long baseline, carried out to investigate the radio structure of the knotty jet in 4C+19.44 at $z$=0.719. The main aim of these low radio frequency observations is to determine the spectral shape at hundreds of MHz to verify the possible presence of spectral curvature. This LOFAR radio image perfectly matches the resolution of those present in the JVLA archive allowing us to perform the desired study. 

Adopting several regions along the radio jet, selected according to the radio-X-ray comparison \citep{xjet,harris17}, we have measured radio flux densities at 159MHz for 11 jet knots in 4C+19.44 ranging between 4 and 17.7 arcsec angular separation form the radio core of the quasar { together with that of the southern hotspot located at 25.7\arcsec from the position of the radio core}. 

Our main results can be summarized as follows.
\begin{itemize}
\item For several jet knots the value of the 159 MHz flux densities lies below the values computed extrapolating the GHz radio spectrum, clearly indicating the presence of spectral curvature as occurs in other source classes.
\item Adopting larger regions along the radio jet we found that the spectral curvature $b$ increases (i.e., narrower spectra) moving far away from the radio core. 
\end{itemize}

We propose to interpret the spectral curvature as due the intrinsic, curved, spectral shape of the emitting particle energy distribution when stochastic acceleration mechanisms occur \citep[see e.g.][for a recent description]{massaro06,tramacere11}. Values of the spectral curvature increasing along the jet are indeed consistent with less efficient acceleration in particles (i.e., electrons) present in knots more distant from the radio core since fractional acceleration gain is inversely proportional to the spectral curvature \citep{tramacere07}. According to previous analyses carried out for different radio extended structures, such as lobes or hotspots, we neglected interpretations of the observed radio spectral curvature as due to absorption plasma effects and synchrotron aging since they failed in previous cases \citep[see e.g,][as previously mentioned]{tingay03,duffy12,mckean16}, favoring a direct link with particle acceleration properties \citep[see also hypothesis made for PKS 1421-490][]{godfrey09}.

To show how the presence of spectral curvature affects the estimate of jet energetics, we have computed the ratios between the magnetic field strength estimated using a simple power-law $vs$ a log-parabolic (i.e. log-normal) distribution as function of both the PED curvature $r$ and the minimum value of the Lorentz factor for the PED: $\gamma_{min}$. Differences between the two models are in the range between a few percent and { up to $\sim$30\%} for values of the spectral curvature observed in the jet knots of 4C+19.44.

LOFAR observations, as those carried out for 4C+19.44 could be also used to improve calculations of the IC/CMB model for which the knowledge of the magnetic field strength and the extrapolation at low energies of the PED is crucial \citep{harris79}.

\begin{acknowledgements}
We thank the anonymous referee for useful comments that led to improvements in the paper presentation.\\
Dan Harris passed away on 2015 December 6th. His career spanned much of the history of radio and X-ray astronomy. His passion, insight, and contributions will always be remembered. \\
LOFAR, the Low Frequency Array designed and constructed by ASTRON, has facilities in several countries, that are owned by various parties (each with their own funding sources), and that are collectively operated by the International LOFAR Telescope (ILT) foundation under a joint scientific policy.
The authors would like to thank the LOFAR observatory staff for their assistance in obtaining and handling of this large data set. \\
This work is supported by the "Departments of Excellence 2018 - 2022" Grant awarded by the Italian Ministry of Education, University and Research (MIUR) (L. 232/2016). This research has made use of resources provided by the Compagnia di San Paolo for the grant awarded on the BLENV project (S1618\_L1\_MASF\_01) and by the Ministry of Education, Universities and Research for the grant MASF\_FFABR\_17\_01. This investigation is supported by the National Aeronautics and Space Administration (NASA) grants GO4-15096X, AR6-17012X and GO6-17081X. A.P. acknowledges financial support from the Consorzio Interuniversitario per la fisica Spaziale (CIFS) under the agreement related to the grant MASF\_CONTR\_FIN\_18\_02.
SGJ acknowledges support from Fermi GI program grant 80NSSC17K0649 and Russian Science Foundation grant 17-12-01029.
he research leading to these results has received funding from the European Research Council under the European Union's Seventh Framework Programme (FP/2007-2013) / ERC Advanced Grant RADIOLIFE-320745
\end{acknowledgements}

\appendix
\section{LOFAR observations and data reduction procedure}

We obtained LOFAR HBA observations on May 14, 2014, from 19:12 to 04:17 UTC on May 15, 2014 under project code  L227113 (PI:D. Harris). The parameters of the observations are summarized in Table \ref{t_obs}. The observation included almost all of the LOFAR high band array (HBA) stations available at the time of the observation, with 23 of the 24 core stations, 13 of 14 remote stations, and all 8 international stations participating.  The core stations observed in HBA\_JOINED mode, where the two `ears' of the station are summed to form a single station.  The international stations consisted of five German stations (DE601, DE602, DE603, DE604, DE605), the one in France (FR606), in Sweden (SE607), and UK (UK608).
This provides baselines lengths ranging from 100~m to 1300~km, although the shortest baselines are of limited utility in the HBA-joined mode.

For this observation we used LOFAR's multi-beaming capabilities. The total available bandwidth was split equally between two simultaneous beams of width 48 MHz (240 sub-bands, each 0.1953 MHz wide), centered on 150 MHz. We placed half of the available instantaneous bandwidth on the target 4C+19.44. The other 244 subbands were pointed towards the nearby calibrator 3C286. 
A few sub-bands were corrupted due to issues with the LOFAR offline storage system; the affected data was flagged, with a negligible effect on sensitivity.
We observed the target for a total on-source time of 9~hrs with integration time of 1~s.  


The first step in the data processing is the automatic flagging of radio frequency interference (RFI) with the AOFlagger \citep{offringa10} and subsequent averaging in time and frequency to 2 sec and 4 channels per subband. This was done with the Observatory pre-processing pipeline and the only pre-processed data was saved to the long term archive (LTA). We then downloaded the archived data and processed it further manually using a modified version of the long baseline pipeline \citep{jackson16}.  After an initial inspection of the data, we flagged the stations  CS001, CS006, CS013, and RS508.  We flagged the data again using the AOFlagger at this reduced resolution, and then finally averaged to 1 channel per subband and 8 second integrations.


The data were initially calibrated with the BlackBoard Selfcal (BBS) software system \citep{pandey09}, using the contemporaneous observation of 3C286. For 3C286 we used the source model from \cite{scaife12} to set the flux scale. We assume that 3C~286 is unpolarized over the observed frequency range and generate full-polarisation solutions on a per-subband basis with a solution interval of 8 seconds.  
Since 3C~286 was observed simultaneously with 4C+19.44 we applied the derived gain and phase solutions directly to 4C+19.44. We estimate the uncertainty on the flux densities of the order of 15 \%.  


However, due to the substantial angular separation between 3C286 and 4C19.44 (over 10 degrees), we expect significant residual calibration errors for the international stations in the direction of the 4C19.44 after the application of these calibration due primarily to the difference in ionospheric path length.  This manifests as a frequency dependent phase change, that can change rapidly with time.  Following the techniques described in  Varenius et al. (2015, 2016), we handle this challenge by calibrating the long baselines using the task FRING in AIPS (Greisen 2003), which solves for a delay and rate.  

We used an iterative procedure of calibration and imaging, beginning with a point source model for 4C19.44.  Only baselines to the international stations were used for imaging, meaning that the visibilities are dominated by 4C19.44 (as other bright and compact sources outside the imaged field are far enough away that time and bandwidth smearing attenuates their contribution).  Pleasingly, the expected morphology was rapidly recovered within a few iterations.  Once an initial model was obtained, we further refined the calibration using amplitude and phase self-calibration (deriving one solution per 16-subband chunk) 
Again, an iterative procedure of calibration and imaging was followed until no further changes to the recovered source image were detectable.  At this point, with the best possible calibration obtained, we were able make images of the target field using a range of $u,v$ lower limits (and hence resolutions) and image sizes.

In effect, by using only baselines to the international stations during the calibration process, we follow an ``outside-in'' method of calibration as was also used successfully for International LOFAR by, e.g, Varenius et al. (2015, 2016).


\begin{table*}
 \centering
  \begin{tabular}{|l|l|l|l|} \hline
  Parameter                & LOFAR HBA    \\ \hline
  Data ID                  & L227113      \\
  Field center RA (J2000)  & 13:57:04.40  \\
  Field center DEC (J2000) & +19:19:07.0  \\
  Observing date           & 2014 May 14  \\
  Total on-source time     & 9 h          \\
  Frequency range          & 115-189 MHz  \\
  Number of sub-bands      & 488          \\
  Width of a sub-band      & 0.195 MHz    \\
  Channels per subband     & 64           \\
  Integration time         & 1 s          \\ \hline
  \end{tabular}
 \caption{Details of the LOFAR observations.}
 \label{t_obs}
\end{table*}

\section{Flux densities of radio knots}

\begin{table}
 \centering
\caption{Radio flux densities extracted from the LOFAR and JVLA images for the radio knots: s2.1,s4.0, s5.3, s6.6, s8.3, s10.0. The knots are indicated following the nomenclature used in Fig.~\ref{fig:jet2}.}
\begin{tabular}{|lll|}
\hline
Region & Frequency & $S_{\nu}$ \\
       & (GHz)     & mJy \\
\hline
\noalign{\smallskip}			         
\hline
  s2.1 & 0.16 & 522.98\\
  s2.1 & 1.36 & 136.31\\
  s2.1 & 1.44 & 135.41\\
  s2.1 & 4.86 & 67.36\\
  s2.1 & 14.96 & 23.68\\
\hline
\noalign{\smallskip}			         
  s4.0 & 0.16 & 99.64\\
  s4.0 & 1.36 & 32.61\\
  s4.0 & 1.44 & 33.13\\
  s4.0 & 4.86 & 14.78\\
  s4.0 & 14.96 & 7.2\\
\hline
\noalign{\smallskip}			         
  s5.3 & 0.16 & 37.87\\
  s5.3 & 1.36 & 15.32\\
  s5.3 & 1.44 & 14.38\\
  s5.3 & 4.86 & 6.89\\
  s5.3 & 14.96 & 2.93\\
\hline
\noalign{\smallskip}			         
  s6.6 & 0.16 & 47.08\\
  s6.6 & 1.36 & 20.4\\
  s6.6 & 1.44 & 20.34\\
  s6.6 & 4.86 & 9.55\\
  s6.6 & 14.96 & 4.56\\
\hline
\noalign{\smallskip}			         
  s8.3 & 0.16 & 84.92\\
  s8.3 & 1.36 & 35.18\\
  s8.3 & 1.44 & 34.94\\
  s8.3 & 4.86 & 16.09\\
  s8.3 & 14.96 & 7.58\\
\hline
\noalign{\smallskip}			         
  s10.0 & 0.16 & 12.83\\
  s10.0 & 1.36 & 12.28\\
  s10.0 & 1.44 & 12.77\\
  s10.0 & 4.86 & 5.98\\
  s10.0 & 14.96 & 2.28\\
\hline
\noalign{\smallskip}			           
\end{tabular}
\label{tab:fluxden1}
\end{table}

\begin{table}
 \centering
\caption{Radio flux densities extracted from the LOFAR and JVLA images for the radio knots: s11.2, s12.9, s14.6, s15.9, s17.7 { and the southern hotspot} s25.7. The knots are indicated following the nomenclature used in Fig.~\ref{fig:jet2}.}
\begin{tabular}{|lll|}
\hline
Region & Frequency & $S_{\nu}$ \\
       & (GHz)     & mJy \\
\hline
\noalign{\smallskip}			         
\hline
  s11.2 & 0.16 & 9.72\\
  s11.2 & 1.36 & 9.83\\
  s11.2 & 1.44 & 10.05\\
  s11.2 & 4.86 & 4.23\\
  s11.2 & 14.96 & 2.03\\
\hline
\noalign{\smallskip}			         
  s12.9 & 0.16 & 15.2\\
  s12.9 & 1.36 & 17.75\\
  s12.9 & 1.44 & 17.83\\
  s12.9 & 4.86 & 7.9\\
  s12.9 & 14.96 & 4.1\\
\hline
\noalign{\smallskip}			         
  s14.6 & 0.16 & 24.2\\
  s14.6 & 1.36 & 15.61\\
  s14.6 & 1.44 & 15.63\\
  s14.6 & 4.86 & 7.04\\
  s14.6 & 14.96 & 3.51\\
\hline
\noalign{\smallskip}			         
  s15.9 & 0.16 & 1.34\\
  s15.9 & 1.36 & 7.07\\
  s15.9 & 1.44 & 7.0\\
  s15.9 & 4.86 & 3.06\\
  s15.9 & 14.96 & 1.1\\
\hline
\noalign{\smallskip}			         
  s17.7 & 0.16 & 4.35\\
  s17.7 & 1.36 & 4.7\\
  s17.7 & 1.44 & 4.75\\
  s17.7 & 4.86 & 1.94\\
  s17.7 & 14.96 & 0.87\\
\hline
\noalign{\smallskip}			         
  s25.7 & 0.16 & 16.86\\
  s25.7 & 1.36 & 6.38\\
  s25.7 & 1.44 & 5.95\\
  s25.7 & 4.86 & 2.31\\
  s25.7 & 14.96 & 1.5\\
\hline
\noalign{\smallskip}			           
\end{tabular}
\label{tab:fluxden2}
\end{table}

\begin{table}
 \centering
\caption{Radio flux densities extracted from the LOFAR and JVLA images for larger regions including more radio knots. The knots are indicated following the nomenclature used in Fig.~\ref{fig:jet2}.}
\begin{tabular}{|lll|}
\hline
Region & Frequency & $S_{\nu}$ \\
       & (GHz)     & mJy \\
\hline
\noalign{\smallskip}			         
\hline
  lower & 0.16 & 50.55\\
  lower & 1.36 & 54.01\\
  lower & 1.44 & 53.94\\
  lower & 4.86 & 23.79\\
  lower & 14.96 & 10.95\\
\hline
\noalign{\smallskip}			         
  middle & 0.16 & 181.1\\
  middle & 1.36 & 89.9\\
  middle & 1.44 & 88.6\\
  middle & 4.86 & 40.71\\
  middle & 14.96 & 17.15\\
\hline
\noalign{\smallskip}			         
  upper & 0.16 & 616.65\\
  upper & 1.36 & 161.08\\
  upper & 1.44 & 160.9\\
  upper & 4.86 & 74.69\\
  upper & 14.96 & 30.39\\
\hline
\noalign{\smallskip}			         
  full jet & 0.16 & 849.41\\
  full jet & 1.36 & 294.37\\
  full jet & 1.44 & 293.83\\
  full jet & 4.86 & 137.56\\
  full jet & 14.96 & 59.08\\
\hline
\noalign{\smallskip}			           
\end{tabular}
\label{tab:fluxden3}
\end{table}
\end{document}